\documentclass{article}

 \usepackage[preprint]{nips_2018}

\usepackage[utf8]{inputenc} 
\usepackage[T1]{fontenc}    
\usepackage{hyperref}       
\usepackage{url}            
\usepackage{booktabs}       
\usepackage{amsfonts}       
\usepackage{nicefrac}       
\usepackage{microtype}      
\usepackage{graphicx}

\usepackage{changepage}

\title{Insect cyborgs: Bio-mimetic feature generators improve machine learning accuracy on limited data}

\author{
  Charles B. ~Delahunt\\  
  Computational Neuroscience Center\\
  University of Washington\\
  Seattle, WA, 98195 \\
  \texttt{delahunt@uw.edu} \\
 \And
  J.~Nathan Kutz\\
  Department of Applied Math\\
  University of Washington\\
  Seattle, WA, 98195 \\
 \texttt{kutz@uw.edu} \\
}

\begin{document}

\maketitle

\begin{abstract}
Machine learning (ML) classifiers typically benefit from more informative input features.
We seek to auto-generate stronger feature sets, to aid ML methods faced with limited training data.
Biological neural nets (BNNs) excel at fast learning, implying that they extract highly informative features. 
In particular, the insect olfactory network  learns new odors very rapidly, by means of three key elements:
A competitive inhibition layer; a high-dimensional sparse plastic layer; and Hebbian updates of synaptic weights.
 
In this work we deploy MothNet, a computational model of the moth olfactory network, as an automatic feature generator.
Attached as a front-end pre-processor, MothNet's readout neurons provide new features, derived from the original features, for use by standard ML classifiers. 
We find that these ``insect cyborgs'' (part BNN and part ML method) have significantly better performance than baseline ML methods alone on vectorized MNIST and Omniglot data sets.
Relative reduction in test set error averages 20\% to 60\%. 

The MothNet feature generator also substantially out-performs other feature generating methods including PCA, PLS, and NNs, as well as pre-training to initialize NN weights. 
These results highlight the potential value of BNN-inspired feature generators in the ML context.
\end{abstract}


\section{Introduction}

Machine learning (ML) methods in general, and neural nets (NNs) with backprop in particular, have posted tremendous successes in recent years \citep{schmidhuber2014, goodfellow2016deep}. 
However, these methods, and NNs in particular, typically require large amounts of training data to attain high performance. 
This creates bottlenecks to deployment, and constrains the types of problems that can be addressed \citep{koller2018}. 
The limited-data constraint is typical of a large and important group of ML targets, including tasks that use medical, scientific, or field-collected data, and also artificial intelligence efforts focused on rapid learning.
We seek to improve ML methods' ability to learn from limited data by improving the input features which an arbitrary ML method can use for training.
In particular, we propose an architecture that can be bolted onto the front end of an ML method, and which automatically generates, from the existing features, a new set of strongly class-separating features  to supplement (or even replace) the existing features.

\begin{figure}[ht]
\begin{center}
\centerline{\includegraphics[width=0.7\columnwidth]{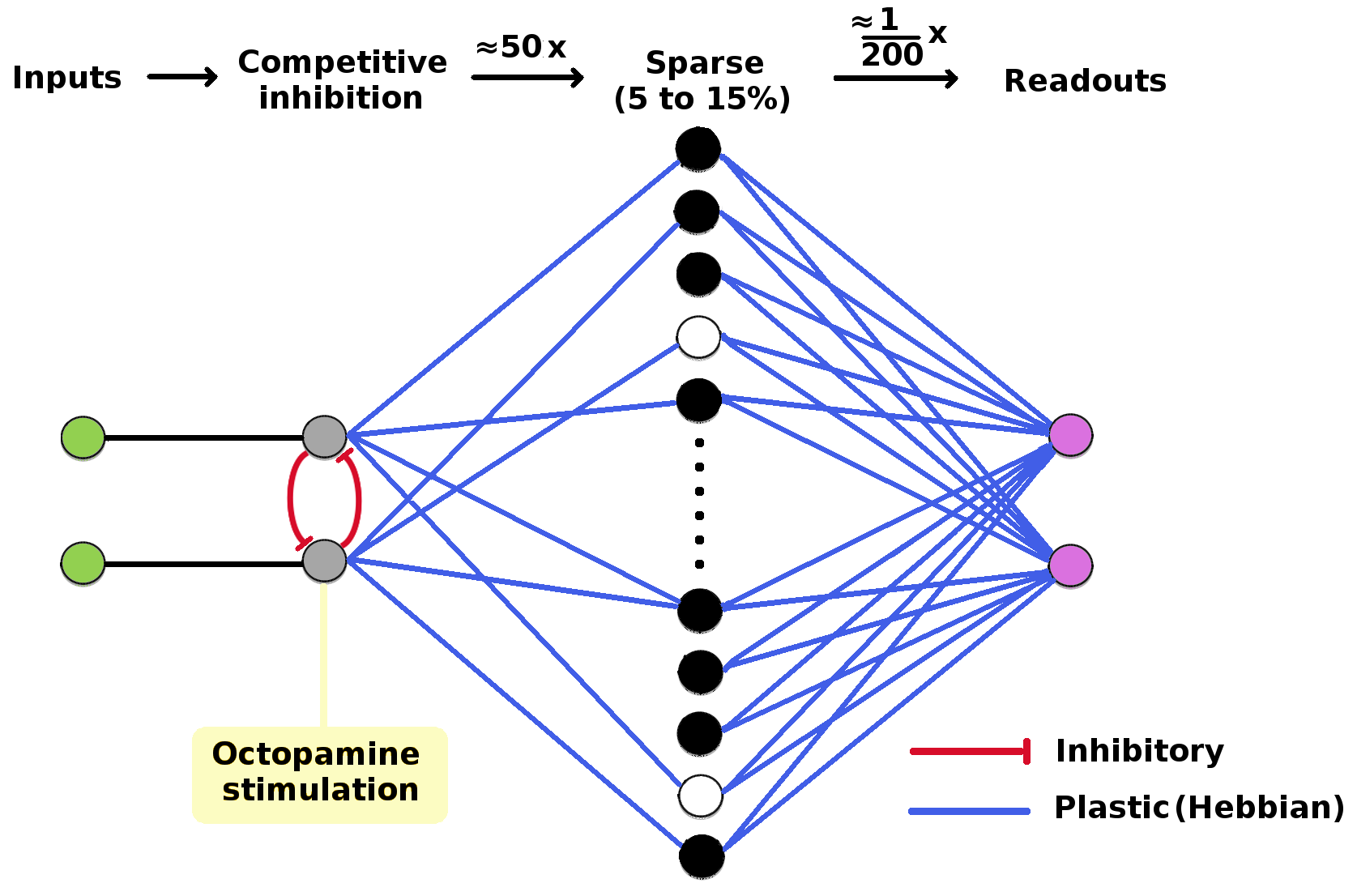}}
  \caption{Schematic of the Moth Olfactory Network. 
  Input features feed 1-to-1 into a 85-unit layer with competitive all-to-all inhibition (the AL). 
  The AL projects with sparse, random connectivity (about 15\%) into a 2500-unit sparsely-firing layer (the MB, with 5\% to 10\% activity).
  The MB projects densely to the Readout Neurons.
  The AL is not plastic.
  The only plastic synaptic weights are those that enter or leave the MB. 
 Training updates are done by Hebbian rule ($\Delta w_{ij} = \alpha f_i f_j$) and unused connections decay towards 0. 
 MothNet instances were generated by randomly assigning connectivity maps and synaptic weights according to template distributions.
  }
  \label{monSchematic} 
\end{center}
\end{figure}

Biological neural nets (BNNs) are able to learn rapidly, even from just one or two samples.
On the assumption that rapid learning requires effective ways to separate classes given limited data, we may look to BNNs for effective feature-generators  \citep{srinivasan2018}.
One of the simplest BNNs that can learn is the insect olfactory network \citep{riffell2013}, containing the Antennal Lobe (AL) \citep{wilson2008} and Mushroom Body(MB) \citep{campbellMushroomBody}, which can learn a new odor given as few as five exposures.
This simple but effective feedforward network is built around three key elements that are ubiquitous in BNN designs: 
Competitive inhibition, high-dimensional sparse layers, and Hebbian update mechanisms.
Specifically, the AL-MB network contains:
(i) A pre-processing layer (the AL) built of units that competitively inhibit each other \citep{bhandawat2007}; 
(ii) Projection, with sparse connectivity and a 10x to 100x dimension shift, up into and then down out of a sparsely-firing high-dimensional layer (the MB) \citep{perisse2013, honeggerTurner2011,ganguli2012}; and
(iii) Hebbian updates of plastic synaptic connections to train the system \citep{hebb, roelfsema2018}. 
Roughly speaking, the Hebbian rule is ``fire together, wire together'', i.e. updates are proportional to the product of firing rates of the sending and receiving neurons, $\Delta w_{ij} = \alpha f_i f_j$. 
Synaptic connections are largely random \citep{caron2013}. 
A schematic is given in Fig \ref{monSchematic}.

MothNet is a computational model of the \textit{Manduca sexta} moth AL-MB \citep{delahuntMoth1}
that demonstrated very rapid learning of vectorized MNIST digits, with performance superior to standard ML methods given $N \leq 10$ training samples per class  \citep{delahuntMnist}. 
 That is, it was able to encode substantial class-relevant information from very few samples.

In this work we examine whether the MothNet architecture can usefully serve, not as a classifier itself, but rather as the first stage of a multi-stage system.
Our goal is to harness its class-information encoding abilities to generate strong features that can improve performance of a   downstream ML classifier. 
In particular, we test the following hypotheses\footnote{See Acknowledgements.}: \newline
$~~~~~$1. The AL-MB architecture has an intrinsic clustering ability, due specifically of the competitive inhibition layer and/or the  sparse high-dimensional layer.
That is, these structures have an inductive bias towards separating classes (just as convolutional NNs have an inductive bias towards distinguishing visual data). \newline
$~~~~~$2. The trained AL-MB is an effective feature generator. 
Its Readout neurons contain class-separating information that will boost an arbitrary ML algorithm's ability to classify test samples.

We tested these hypotheses by combining MothNet with a downstream ML module, so that the Readouts of the trained AL-MB model fed into the ML module as additional features (from the ML perspective, the AL-MB acted as an automatic feature generator; from the biological perspective, the ML module stood in for the downstream processing in more complex BNNs). 
Our Test Cases were (a) a non-spatial, 85-feature, 10-class task derived from the downsampled, vectorized MNIST data set (hereafter ``$v$MNIST'' to emphasize its vectorized, non-spatial, structure); and (b) a non-spatial, 120-feature, 10-class task derived from the downsampled, vectorized Omniglot data set ($v$Omniglot).
We restricted training set size to $N \leq 100 $ samples per class, so that the ML methods did not attain full accuracy on the task using the 85 (or 120) features alone.

We found evidence that these hypotheses are correct: 
The high-dimensional sparse layer and the competitive inhibition layer, in combination with a Hebbian update rule, significantly improve the abilities of ML methods (NN, SVM, and Nearest Neighbors) to classify the test set in all cases.
That is,   the original input features (pixels) contain class-relevant information which is not accessed by the ML methods, but which the MothNet module encodes in a form that is accessible to the ML methods.
If the learning performance of BNNs is any guide, these layers are simple, general-purpose feature generators that can potentially improve performance of ML methods in  tasks where training data is limited.

In addition, the MothNet-generated features significantly out-performed  features generated by PCA (Principal Components Analysis), PLS (Projection to Latent Structures), and NNs, in terms of their ability to improve ML accuracy.
They also out-performed NNs with weights initialized by pre-training on the Omniglot data set.
These results indicate that the insect-derived network generated significantly stronger features than these other  methods.

%
%
 
\section{Experimental setup}
To generate $v$MNIST, we downsampled and preprocessed the MNIST data set \citep{leCunMnist, murphyML} to give samples with 85 pixels-as-features stripped of spatial information, as in \citet{delahuntMnist}.
We note that $v$MNIST is not the ``MNIST data set" considered in its usual context of a task with spatial structure and large pools of training data.
Rather, here the MNIST data served as raw material for a generic non-spatial Test Case.
$v$MNIST had the advantage that our baseline ML methods (Nearest Neighbors, SVM, and Neural Net) did not attain full accuracy at low \textit{N}.
So it acted as a good test of whether the AL-MB can improve classification by ML methods.
The Omniglot dataset \citep{omniglot} is a collection of hand-drawn characters similar to MNIST, except that it contains 1100 classes with only 20 samples each.
A downsampled, non-spatial version was created in similar fashion.

Full network  architecture details of the AL-MB model (MothNet) are given in \citet{delahuntMoth1} and \citet{delahuntMnist}.
Full Matlab code for these cyborg experiments including comparison methods, as well as code for MothNet simulations, can be found at \url{https://github.com/charlesDelahunt/PuttingABugInML}.

MothNet instances were generated randomly from templates.
Key aspects of the MothNet model include: 
(i) Competitive inhibition in the AL. Each neural unit in the AL receives input from one feature, and has two outputs, an inhibitory signal to other neural units in the AL,  and an excitatory signal to the MB.
Thus, each feature tries to dampen other features' presence in the sample's output signature from the AL. %
(ii) Sparsity in the MB, of two types. First, the projections from the AL to the MB are non-dense ($\approx$15\% non-zero);
second, MB neurons fire sparsely in the sense that only the strongest 5\% to 15\% of the total population are allowed to fire (through a mechanism of global inhibition). 
(iii) All weights are non-negative, and are initialized randomly. 
(iv) Weight updates affect only MB$\rightarrow$Readout connections (the AL is not plastic, and AL$\rightarrow$MB learning rates are  slow).
Hebbian updates occur according to: $\Delta w_{ij} = \alpha f_i f_j$ if $f_i f_j > 0$ (growth), and $\Delta w_{ij} = -\delta w_{ij}$ if $f_i f_j = 0$ (decay), where $f_i, f_j$ are two neural firing rates and $w_{ij}$ is their connection weight.

ML methods were treated as follows: 
Nearest-Neighbors and SVM used Matlab built-in functions.
The Neural Nets used Matlab's NN toolbox, with one layer (more layers did not help) and as many hidden units as features (e.g. 85 or 95 for $v$MNIST; more units did not help). 
All hyperparameter details can be found in the online codebase.
We note that our goal was to see if the MothNet-generated features improved on the baseline accuracy of the ML methods  whatever that baseline was.
So the exact ML method hyperparameters were not central, as long as they were reasonable. 
We varied the trained accuracy of baseline methods  by restricting training data.\newline
We ran four sets of experiments:

\subsection{Cyborg vs baseline ML methods on  $v$MNIST }\label{cyborgExperimentSetup}
To assess the benefit of MothNet features, experiments were structured as follows:\newline
$~~~~~$1. A random set of \textit{N} training samples per class were  drawn from  $v$MNIST. \newline
$~~~~~$2. The ML methods trained on these samples, to provide a baseline (switch $\overline{B}$ in Fig \ref{configurationSchematic}).  \newline
$~~~~~$3.  MothNet was trained on these same samples, using time-evolved stochastic differential equation simulations and Hebbian updates (switch $A$ in Fig \ref{configurationSchematic}). \newline
$~~~~~$4. The ML methods were then retrained from scratch, with the Readout Neuron outputs from the trained MothNet instance fed in as additional features (switches $A B$ in Fig \ref{configurationSchematic}). 
These were the ``insect cyborgs'', i.e. an AL-MB feature generator joined to a ML classifier. \newline
$~~~~~$5.  Trained ML accuracy of the baselines and cyborgs were compared to assess the value of  the AL-MB as a feature generator.
These experiments were repeated 13 times per $N$, for each ML method.

\subsection{Cyborg vs baseline ML methods on $v$Omniglot}
These experiments were set up as in (1), but used the $v$Omniglot data set.
For each run, 10 Omniglot classes were randomly chosen.
Thumbnails were first pre-centered, then cropped and subsampled down to 120 pixels and vectorized.
We set $N \leq 15$ to ensure at least 5 test samples per class. 

\subsection{MothNet features vs other feature generators }
To compare the effectiveness of  MothNet features vs  features generated by conventional ML methods, we ran $v$MNIST experiments structured as the MothNet experiments in (1) above, but with the MothNet feature module replaced by one of the following options:\newline
$~~~~~$1.  PCA (Principal Components Analysis) applied to the $v$MNIST training samples.  
The new features were the projections onto each of the top 10 modes. \newline
$~~~~~$2. PLS (Projection to Latent Structures) applied to the $v$MNIST training samples. 
The new features were the projections onto each of the top 10 modes.
We expected PLS to do better than PCA because PLS incorporates class information.  \newline
$~~~~~$3.  NN pre-trained on the $v$MNIST training samples. 
The new features were the (logs of the) 10 output units. 
This feature generator was used as a front end to SVM and Nearest Neighbors only.
Since $v$MNIST has no spatial content, CNNs were not used. \newline
$~~~~~$4. NN with weights initialized by training on an 85-feature $v$Omniglot data set, then trained on the $v$MNIST data as usual (transfer learning). 
This was applied to the NN baseline only. 

\subsection{Relative importance of AL vs MB} 
\begin{figure}[t]
\begin{center}
\centerline{\includegraphics[width=0.8\columnwidth]{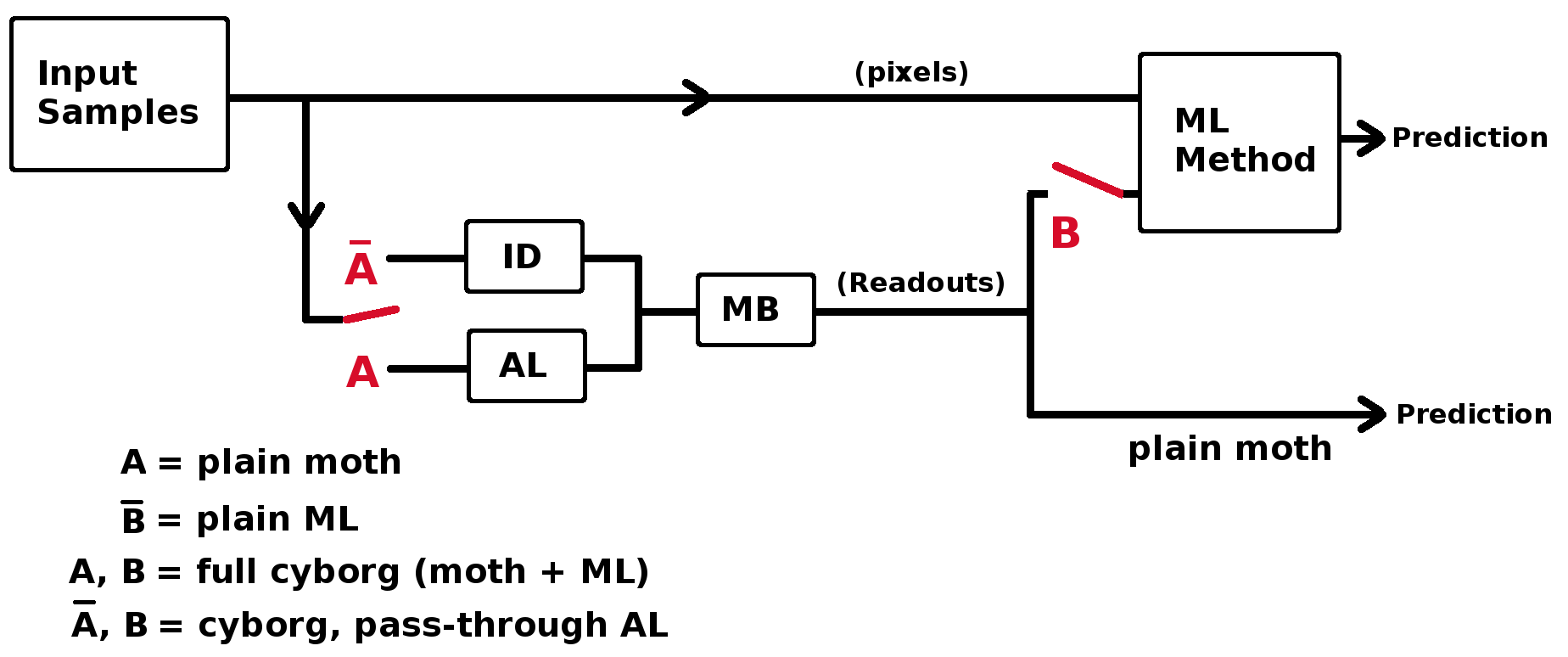}} 
  \caption{Schematic of the various Learner configurations.
 Two switches created the various models. 
 In the ordinary MothNet, input pixels passed through the AL (switch $A$), then the MB, and prediction was based on a log-likelihood over the Readout Neurons as in \citet{delahuntMnist}.
 The ordinary (baseline) ML module accepted only input pixels as features (switch $\overline{B}$). 
 Two cyborg variants were tested:
 In the full cyborg, the Readouts of an ordinary trained MothNet are fed into the ML module as additional features (switches $A B$).
 In a mutant cyborg used to test the role of the AL, Readouts from a trained MothNet with disabled (pass-through) AL fed into the ML module as additional features (switches $\overline{A} B$).
  }
  \label{configurationSchematic}
\end{center}
\end{figure}

The AL-MB has two key structural components, a competitive inhibition layer (the AL) and projection into a high-dimensional sparse layer (the MB) with Hebbian synaptic updates.
These two structures can be deployed separately or together.
In particular, the (trainable) high-dimensional sparse layer can be deployed with or without the competitive inhibition layer.
To assess the relative value of  the competitive inhibition layer, ``mutant'' MothNets were generated from templates that had a pass-through AL  with no lateral inhibition (switch $\overline{A}$ in Fig \ref{configurationSchematic}). 
 Steps 1 to 4 above were followed using these mutant MothNets (so Step 4 corresponded to switches $\overline{A} B$ in Fig \ref{configurationSchematic}). 
The results from step (4) were then compared to those of full cyborgs on $v$MNIST data.

%
%

\section{Results}
\label{results}

On both $v$MNIST and $v$Omniglot, the use of MothNet readouts as features significantly improved accuracy of ML methods, demonstrating that the MothNet architecture effectively captured class-relevant features.
The MothNet-generated features were also far more effective than the comparison feature generators (PCA, PLS, and NN). 

We note that MothNet excels at learning from few samples.
The trained MothNet learners alone, attained a mean accuracy of 65\% to 75\% on $v$MNIST and 80\% to 90\% on $v$Omniglot (given pre-centered starting thumbnails), depending on number of  training samples per class \textit{N}.
However, MothNet features significantly improved ML methods' performance even when the ML methods' baseline accuracies were already greater than MothNet accuracy (e.g. NNs with $N \geq 40$, on $v$MNIST).

Results are given below for each experiment.

%
%

\subsection{Gains due to MothNet features on $v$MNIST}  
MothNet-ML cyborgs, i.e. networks in which the 10 Readouts of the trained MothNet were fed into the ML module as 10 additional features, showed consistently improved test set performance versus their ML baselines on $v$MNIST, for all ML methods at almost all $N > 3$.

\begin{figure*}[t]
\begin{center}
\centerline{\includegraphics[width=0.95\textwidth]{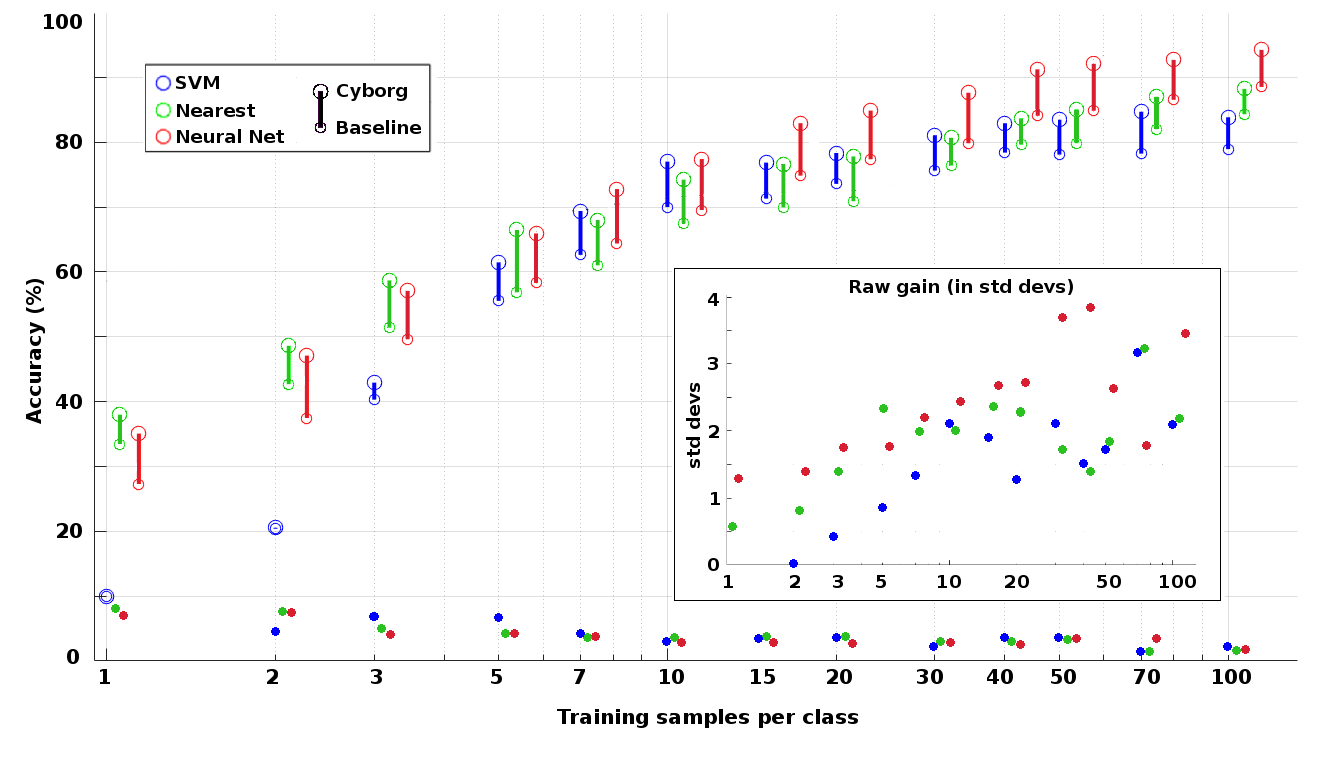}}  
  \caption{Trained accuracy of baseline ML and cyborg classifiers, vs $N$ training samples per class.
  Baseline ML accuracies are shown as small circles, 
  cyborg accuracies are shown as larger circles, and thick vertical bars mark the increase in accuracy.
 Baseline methods'  std dev ($\sigma$) are given as solid dots near the x-axis. 
 Inset: The raw gain in accuracy (cyborg over ML baseline) in units of std dev. 
 MothNet features significantly improved ML accuracy (see Table \ref{cyborgSignificanceTableMnist}).
  }
  \label{comparisonPlainVsCyborg}
\end{center}
\end{figure*}

The ML baseline methods (no added features) started at 10\% to 30\% accuracy for $N$ = 1 sample per class,  and rose to 80\% to 88\% accuracy (depending on method) at \textit{N}=100, where we stopped our sweep.
This baseline accuracy is marked by the lower colored circles in Fig \ref{comparisonPlainVsCyborg}. 
Cyborg accuracy is marked by the upper colored circles in Fig \ref{comparisonPlainVsCyborg}, and the raw gains in accuracy are marked by thick vertical bars.

\begin{figure*}[t]
\begin{center}
\centerline{\includegraphics[width=0.95\linewidth]{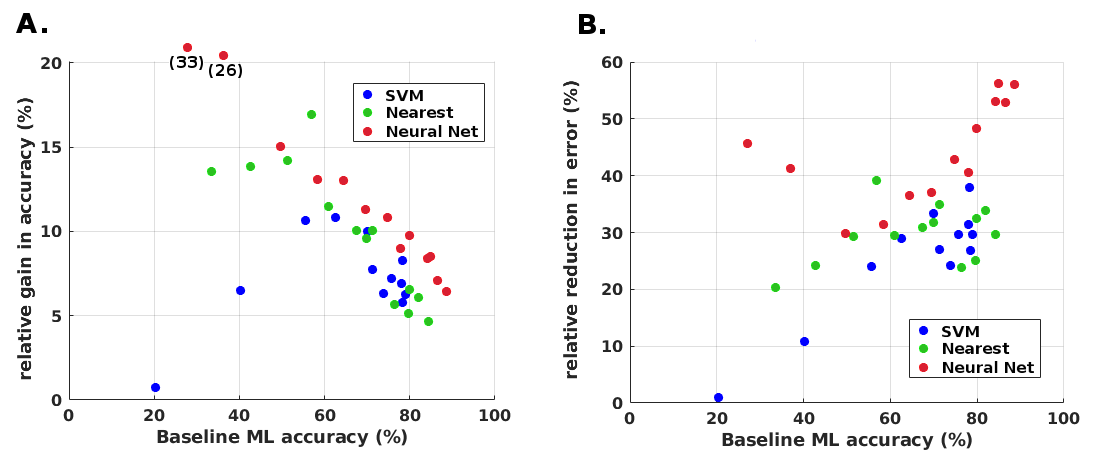}} 
  \caption{ Relative gains in test set performance on $v$MNIST due to MothNet features, vs baseline ML accuracy. 
  {\bf{A:}} Mean relative gains in accuracy over ML baseline. 
{\bf{B:}} Mean relative reduction in test set error was consistently high, especially for NNs. 
  }
  \label{percentGainInAccuracy}
\end{center}
\end{figure*}

Raw increases in accuracy due to MothNet features were fairly stable for all ML models. 
This led to two trends in terms of relative changes.
Relative gains, i.e. as percentage of baseline, were highest at low $N$ training samples per class: 
Average relative gains were 10\% to 33\% at $N \leq 10$, and 6\% to 10\% for $N > 10$ (see Fig \ref{percentGainInAccuracy} A).
Conversely, the relative reduction in test set error, as a percentage of baseline error, increased substantially as baseline accuracy  grew (see Fig \ref{percentGainInAccuracy} B).
Thus, MothNet cyborgs reduced test set error by over 50\% on the most accurate models, such as NNs with $>$80\% baseline accuracy.
Of the ML methods, Neural Net cyborgs had the best performance and posted the largest gains.

 \begin{table}[b]
  \begin{center}
    \caption{\textit{P}-values of gains in accuracy due to MothNet features on $v$MNIST,  for each ML baseline method at each $N$. 
We set the baseline accuracy to be the null hypothesis $\mathcal{N}(\mu_1, \sigma_1)$; $\mu_2$ to the mean accuracy using MothNet features; and calculate $P(x \geq \mu_2 ~| ~\mu_1, \sigma_1, n)$ using Student's, with $n$ = 13 runs.  
``0*'' means \textit{P}-value $< 0.0005$.    }
\label{cyborgSignificanceTableMnist}
          \begin{tabular}{l|c|c|c|c|c|c|c|c|c|c|c|c}     
      \textbf{Method} & \textit{N} =1 & 2 & 3 & 5 & 7 & 10 &  15 & 20 & 30 & 50 & 70 & 100\\ 
      \hline
      NearNeigh  &   0.032 &  0.007 & 0*  &   0* &   0* & 0* &  0*  &   0*  &  0*  &  0*&  0*  & 0*    \\
      SVM              &   NA & 0.413& 0*  &  0.004 &   0* & 0* &  0*  &   0*  &  0*  &  0*&  0*  & 0*    \\
      Neural Net &  0.293  &  0.152& 0.029 &   0* &   0* & 0* &  0*  &   0*  &  0*  &  0*&  0*  & 0*   \\
    \end{tabular}
  \end{center}
\end{table}
 

Gains were strong for all $N \geq 3$, being on average one to 4 std devs (\textit{i.e.} Mahalanobis distance), seen  in the inset of Fig \ref{comparisonPlainVsCyborg}. 
 \textit{P}-values were almost all less than $10^{-3}$. 
Table \ref{cyborgSignificanceTableMnist} gives the \textit{P}-values of the gains due to MothNet features, for each $N$ and ML method.

Remarkably, adding a MothNet front-end improved ML accuracy even in cases where the ML module baseline already exceeded the accuracy ceiling of MothNet,   at $N$ = 15 to 100 samples per class.
This  implies that the Readouts of MothNet contain valuable clustering information which ML methods  can leverage more effectively than MothNet itself does.


%
%

\subsection{Gains due to MothNet features on $v$Omniglot}
 
\begin{figure*}[h]
\begin{center}
\centerline{\includegraphics[width=1\linewidth]{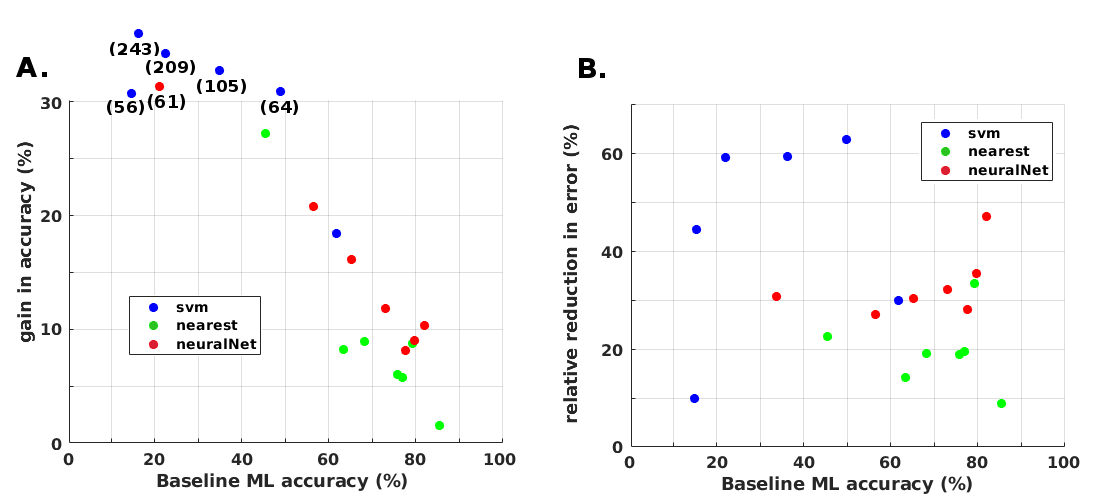}} 
  \caption{ Effects on test set accuracy of cyborgs (using only MothNet readouts as features) over baseline ML, 
   on $v$Omniglot.  
  {\bf{A:}} Mean relative gains in accuracy over ML baseline, due to MothNet features. 
  SVMs benefitted greatly, NNs benefitted with moderate significance ($p < 0.2$), and Nearest Neighbors did not benefit significantly.
{\bf{B:}} Relative reduction in test set error due to MothNet features was roughly 20\% for Nearest Neighbors, 60\% for SVM, and 30\% for NNs. 
  }
  \label{omniglotGainsReductions}
\end{center}
\end{figure*}

As in the $v$MNIST case, MothNet cyborgs posted high relative gains in accuracy vs baseline ML methods on $v$Omniglot data.
Remarkably,  the highest gains resulted from using only the MothNet outputs as features and ignoring the original image pixels, a result that underscores the effectiveness of the MothNet-generated features.
We note that MothNet is a rapid learner (test set accuracy was 81\% to 88\%), and ML methods are weakest at low $N$:
baseline ML accuracies were 45\% to 85\% (Nearest Neighbor), 15\% to 62\% (SVM), and 34\% to 82\% (NN).
So the advantage of solely using MothNet features made sense on this data set since by definition $N \leq 15$. 
The relative increase in accuracy and reduction in error due to AL-MB features are shown in Figure \ref{omniglotGainsReductions}.


\begin{table}[h]
  \begin{center}
    \caption{\textit{P}-values of gains in accuracy  due to MothNet features on $v$Omniglot, for each ML baseline method at each $N$. 
We set the ML baseline accuracy as null hypothesis $\mathcal{N}(\mu_1, \sigma_1)$; $\mu_2$ to the mean accuracy using only MothNet features (i.e. ignoring image pixels); and calculate $P(\mu_2~ | ~\mu_1, \sigma_1, n)$  using Student's, with $n$ = 17 runs.
``0*'' means \textit{P}-value $< 0.0005$.    } 
\label{cyborgSignificanceTable}
    \begin{tabular}{l|c|c|c|c|c|c|c|c|c|c|c|c|c}  
      \textbf{Method} & \textit{N} =1 & 2 & 3 & 5 & 7 & 10 &  15 \\ 
      \hline
      NearNeigh  &  0*      &   0.003   &   0.006    &  0.014     &  0.003    &   0.001  & 0.28   \\
      SVM               & NA    &    0*    &   0*     &  0*      &  0*     &   0*   & 0*    \\
      Neural Net   &  0*     &   0*    &   0*     &  0*      &  0*     &   0*   & 0*    \\
    \end{tabular}
  \end{center}
\end{table}

MothNet-generated features resulted in high relative gains in accuracy ($\approx$ 5\% to 20\%, as well as several much larger gains).
However, due to low $N$ ($\leq15$),  std dev of baseline ML accuracy was always high ($ \approx 10\% $, roughly  double that of $v$MNIST). 
Thus, for Nearest Neighbor and Neural Net the gains were not as strongly significant in a $p$-value sense, as shown in Table \ref{cyborgSignificanceTable}.
Cyborgs increased Nearest Neighbor accuracies by around 0.7 std devs, and Neural Net accuracy by about 1.2 to 1.5 std devs, corresponding to $p$-values $\approx$ 0.15 to 0.2 (expressed as 15 to 20 in Table \ref{cyborgSignificanceTable}).
SVMs had poor baseline performance on $v$Omniglot and posted large gains by using MothNet features (Figure \ref{omniglotGainsReductions}).

%
%

\subsection{Comparison to other feature generators} 
To compare with other automated feature generating methods, we ran the cyborg framework on $v$MNIST using PCA (Principal Components Analysis, projections onto top 10 modes), PLS (Projection to Latent Structures, projection onto top 10 modes), and NN (logs of the 10 output units). 
In each case, the method used the training samples to generate 10 new features. 
Each method was run using a Matlab built-in function. 

With few exceptions, MothNet features were far more effective than the other methods. 
Tables \ref{otherMethodsOnNearestNeighbor}, \ref{otherMethodsOnSvm}, and \ref{otherMethodsOnNeuralNet} give, for each  ML classifier, the relative increase  in mean accuracy due to the various feature generators (or to pre-training).  
``MothNet'' refers to using MothNet features.
 13 runs per data point.

\begin{table}[h]
  \begin{center}
    \caption{Nearest Neighbor: Mean relative percentage increase in accuracy over Nearest Neighbor as baseline classifier, due to various feature generators (``F Gen'').   }
\label{otherMethodsOnNearestNeighbor}
    \begin{tabular}{l|c|c|c|c|c|c|c|c|c|c|c|c} 
      \textbf{F Gen} & \textit{N}=1 & 2 & 3 & 5 & 7 & 10 &  15 & 20 & 30 & 50 & 70 & 100\\ 
      \hline
      PCA              & -67     &  0.7 &  0.6  & 1.4     & 1.2     &  1.2  &  1.5  & 1     & 1.3    & 0.0   &  0.9  & 1.5   \\
      PLS               & NA   & 1.4  & 0.6   &  1.6  &  2.1  & 1.5  &  1.1  &  1.9 &   1.2  &  0.4   &  0.9  &   -0.1   \\
      NN &  -1.4 &   1.3 & 2.1  & 1.5  &  2.6  & 2.1   & 4.4   & 3.2  &  4.7   & 3.9  &   3.9  &   3.7   \\
    MothNet    &  \textbf{13.6}  &    \textbf{13.9} &	  \textbf{14.2} 	&  \textbf{16.9} &    \textbf{11.5}&    \textbf{10}  &     \textbf{9.6}  &    \textbf{10}  &     \textbf{5.6}  &     \textbf{ 6.6}   &    \textbf{6.1}  &     \textbf{4.7}   \\
      
    \end{tabular}
  \end{center}
  \begin{center}
    \caption{SVM: Mean relative percentage increase in accuracy over SVM as baseline classifier, due to various feature generators (``F Gen'').   }
\label{otherMethodsOnSvm}
          \begin{tabular}{l|c|c|c|c|c|c|c|c|c|c|c|c } 
      \textbf{F Gen} & \textit{N}=1 & 2 & 3 & 5 & 7 & 10 &  15 & 20 & 30 & 50 & 70 & 100\\ 
      \hline
       PCA              &NA	&   \textbf{12.2}  & -0.4 &	-1.4 &  0.3 &  0.2   & 0.2   & -0.9  & 0.3   &  -0.8   & -1.4  &  -0.5   \\
      PLS               & NA   &-14 & 4.2  	&3.5 &   1.5  & -0.2  & -2.6  & -4   &  -5.4   & -5.3  &  -5.1  &  -5.5   \\
      NN &  NA   &6.8  &	-1.3 	&-3.7  & -2   & -0.9  & 1.7  &  0.5  &  4.3 &  4.1   &  4.9   &  4.9   \\
      MothNet    &   NA  & 0.8  &	  \textbf{6.5}  &	  \textbf{10.7}  &   \textbf{11 } &  \textbf{10}   &    \textbf{7.8}  &    \textbf{6.3}  &    \textbf{7.2}   &   \textbf{ 6.9}   &    \textbf{8.3}  &     \textbf{6.2}   \\  
      
    \end{tabular}
  \end{center}
  \begin{center}
    \caption{Neural Net: Mean relative percentage increase in accuracy over NN as baseline classifier, due to various feature generators (``F Gen'').
    ``preTrain '' means: Initialize NN weights by training on Omniglot, then train  on the $v$MNIST data. }
\label{otherMethodsOnNeuralNet}

          \begin{tabular}{l|c|c|c|c|c|c|c|c|c|c|c|c} 
      \textbf{F Gen} & \textit{N}=1 & 2 & 3 & 5 & 7 & 10 &  15 & 20 & 30 & 50 & 70 & 100\\ 
      \hline
       PCA              &-57&	0.2	&-0.8	&1.2		&2.6   &	1.7    &	0.3	&	1.3	&	-0.3 &  0.2  &  0.3 &   0.2  \\
      PLS               & NA	&0.2	&5.9	&	1.0	&	1.5  & 	2.8    &	-0.2&	1.2	&	0.3    &   1.6  &  1.5  &  1.9   \\
      preTrain &    \textbf{15}	&4.2&	5.8	&	-3.1&	-1.1  &	0.2    &	1.3		&1.5		&-3.4  &   -0.4   &-4.7   &-1.1   \\
      MothNet &   4	&  \textbf{17}	&  \textbf{15}	&	  \textbf{13.1}	&  \textbf{13}  &  	  \textbf{11.3}   &	  \textbf{10.8}	&  \textbf{9.0}	&	  \textbf{9.7}    &    \textbf{ 8.5} &    \textbf{ 7.1}  &    \textbf{6.4}   \\
      
    \end{tabular}
  \end{center}
\end{table}
%
%

\subsection{Relative contribution of the AL and MB layers}

MothNet has two key structures, a competitive inhibition layer (the AL) and a high-dimensional, sparse layer (the MB).
Cyborgs built from MothNets with a pass-through (identity) AL still posted significant improvements in accuracy over baseline ML methods. 
The gains of cyborgs with pass-through ALs were generally between 60\% and 100\% of the gains posted by cyborgs with normal ALs (see Table \ref{mbValueTable}),  
suggesting that the high-dimensional, trainable layer (the MB) was of primary importance. 
However, the competitive inhibition of the AL layer clearly added value in terms of generating strong features, contributing up to 40\% of the total gain.
NNs benefitted most from the competitive inhibition layer.

In terms of overall effect on downstream ML modules, the AL enabled slightly better, more reliable gains: 
Averaged over all ML methods and all numbers of training samples $N$, a functioning AL gave mean raw increase in accuracy = 5.6\%, standard error $(\frac{\sigma}{\mu})$ = 0.38; while a pass-through AL gave mean raw increase in accuracy = 5.0\%, standard error = 0.43. ($v$MNIST, 13 runs per data point.) 
  
 
\begin{table}[h]
  \begin{center}
    \caption{Relative importance of the MB, vs number of training samples per class $N$. 
Entries give the gains posted by cyborgs with pass-through ALs as a percentage of the gains of full cyborgs (shown in Fig \ref{percentGainInAccuracy}), for the three ML methods. 
Entries = 100\% indicate that average gains from the pass-through AL were $\geq$ average gains from the normal AL. }
\label{mbValueTable}
    \begin{tabular}{l|c|c|c|c|c|c|c|c|c|c|c|c|c} 
      \textbf{Method} & \textit{N} =1 & 2 & 3 & 5 & 7 & 10 &  15 & 20 & 30 & 40 & 50 & 70 & 100\\ 
      \hline
      NearNeigh  & 82    &  100&  91    & 76     &  100   &  100  &  58    & 74     & 88    &  64    & 100   & 100  & 65   \\
      SVM              & NA   &NA   & 100   &  87   &  79      & 97      & 75     &  94   &  98     &  82    &  100  & 76   & 15   \\
      NN &  100 & 60   & 62     &  67    &   75    &  91     &  100  &  93    &  100  &  100  &   100 &  82  & 65   \\
    \end{tabular}
  \end{center}
\end{table}
 
  


\section{Discussion}
Strong, automatically-generated feature sets enhance the power of ML algorithms to extract structure from data.
They are always desirable tools, but especially so when training data is limited.
Many ML targets, such as tasks for which data must be manually collected in medical, scientific, or field settings, do not have the luxury of vast amounts of (e.g. internet-generated) training data, so they must extract maximum value from the limited amount available.
This  large class of ML targets  also includes Artificial Intelligence systems that seek adaptive and rapid learning skills.
In this context, biological structures and mechanisms are potentially useful tools, given that BNNs excel at rapid learning.

Our experiments deployed an architecture based on a very simple BNN, the moth olfactory network, to generate features to support ML classifiers. 
The  three key elements of this network  are novel in the context of engineered NNs, but are endemic in BNNs of all complexity levels: (i) a competitive inhibition layer; (ii) a high-dimensional sparse layer; and (iii) a Hebbian plasticity mechanism for weight updates in training.
Our experiments indicate that these structures, as combined in the MothNet model of the insect olfactory network, create a highly effective feature generator whose Readout Neurons contain strong class-specific information.

In particular, using MothNet as a feature generator upstream of standard ML methods consistently improved their learning abilities on both $v$MNIST and $v$Omniglot.
That is, some class-relevant information in the raw feature distributions was not extracted by the ML methods alone, but pre-processing by MothNet made that information accessible. 
In some cases, the ML methods made better use of the MothNet features than MothNet itself. 
For example, relative reduction in test set error exceeded 50\% for NN models with higher ($> 80\%$) baseline accuracies than MothNet on $v$MNIST.

In addition, MothNet features were consistently far more useful than features generated by standard methods such as PCA, PLS, or NNs, and also more useful than pre-training NNs on similar data.
 
These gains in accuracy can be viewed as savings on training data needs: 
For example, with $N$ = 30 training samples per class, a MothNet+NN cyborg attains the same test accuracy (79\%) as a NN baseline attains with $N$ = 100, a savings of over 3x in training data ($v$MNIST).
These savings in training data can be seen in Fig \ref{comparisonPlainVsCyborg} by drawing horizontal lines between cyborg and baseline accuracies.
Savings consistently ranged from 1.5x to 3x.
If these accuracy gains and commensurate savings held for higher numbers of training samples in more difficult tasks, the savings in data requirements would be substantial, an important benefit for many ML use-cases.

Not only can the structures found in the AL-MB be readily prepended as feature generators to arbitrary ML modules, as we did here, but they can perhaps also be inserted as layers into deep NNs. 
Indeed, this is what BNNs appear to do. 

\subsection{Comparison of the Mushroom Body to sparse autoencoders}
The insect MB is a biological means to project codes into a sparse, high-dimensional space.
It naturally  brings to mind sparse autoencoders (SAs) \citep{ngSparseAutoencoder, makhzani}.
However, there are several differences, beyond the fact that MBs are not trying to match the identity function.

First, in SAs the goal is typically to detect lower-dimensional structures that carry the input data. 
Thus the sparse layers of SAs have fewer active neurons than the nominal dimension of the input. 
In the MB, the number of neurons increases manyfold (e.g. 50x), so that even with enforced sparsity the number of active MB neurons is much greater than the input  dimension: In MothNet there are approximately 150 - 200 active neurons in the MB vs 85 input features. 
The functional effects are also different:
In MNIST experiments in \citet{makhzani}, a sparse layer with 100 active neurons (vs 784 input pixels, i.e. ratio 1:8) captured only very local features and was not effective for feeding into shallow NNs (though it was useful for deeper nets). 
In our experiments, a ratio of 2:1 (i.e. 16x that of the SA) generated  features that were very effective as input to a shallow net. 

Second, there is no off-line training or pre-tuning step, as used in some SAs, though of course Mother Nature has been tinkering with this system for a long time. 
Third,  SAs typically (to our knowledge) require large amounts of training data (e.g. 5000 per class in \citet{makhzani}), while the MB needs as few as one training sample per class to bake in structure that improves classification.
Fourth, the updates in SAs are by backprop, while those in MBs are Hebbian. 
While the ramifications of this difference in update method are unclear, we suspect that the  dissimilarity of the optimizers (MothNet vs ML) was an asset in our experiments.

The MB shares with Reservoir Networks \citep{schrauwen2007} a (non-linear) projection into a high-dimensional space and  projection out to a Readout layer, though this second projection differs by being linear in Reservoir Networks.
A major difference between the MB and a Reservoir Network is that in the MB neurons are not recurrently connected, while in a Reservoir Network they are.

\subsection{Role of the competitive inhibition layer}
The competitive inhibition layer may enhance classification by creating several attractor basins for inputs, each focused according to which subsets of features present most strongly, which in turn depends on the classes.
This might serve to push otherwise similar samples (of different classes) away from each other, towards their respective class attractors, increasing the effective distance between the samples. 
Thus the  outputs of the AL, after this competitive inhibition, may have better separation by class.  

However, in our experiments  on this particular data set, while the competitive inhibition layer (AL) did benefit the downstream ML classifier, it was less important than the sparse layer (MB).
We see two reasons why this might so.
First, the AL has other jobs to do in the insect olfactory network, such as gain control and corraling inputs from the noisy antennae \citep{martin2011, olsenWilson2010}. 
Perhaps these are the AL's primary tasks, and separating input signals is a secondary task.
Second, the MothNet model was transferred to the $v$MNIST task from a model developed to study odor learning that was calibrated to \textit{in vivo} moth data \citep{delahuntMoth1}. 
Perhaps the AL has a larger role in the natural, odor-processing setting, and its transfer to the $v$MNIST task modified the overall balance of the AL-MB system and reduced the importance of the AL relative to the MB.
That said, the best results and also most consistent improvements were posted by generating features using the full AL-MB network.

\subsection{Role of Hebbian updates} 
We suspect that much of the success of BNNs (and MothNet) is due to the Hebbian update mechanism, which appears to be quite distinct from typical ML weight update methods. 
It has no objective function or output-based loss that is pushed back through the network  as in backprop or agent-based reinforcement learning (there is no ``agent'' in the MothNet system).
Hebbian weight updates, either growth or decay, occur on a local ``use it or lose it'' basis.

We also suspect that part of the success of the cyborgs was due to the stacking of two distinct update methods,  Hebbian and backprop.
In our experience, stacking dissimilar ML methods is more productive than stacking similar methods.
This may be one reason MothNet cyborgs delivered improvement to ML accuracy even in cases where the baseline ML accuracy already exceeded the MothNet's top performance: 
each system brings unique structure-extracting skills to the data.
It may also explain why projecting into the high-dimensional MB is not redundant when paired with an SVM, which also projects into a high-dimensional space: The two methods of learning the projections are different.

\subsection{Limitations}
A  practical limitation of this method, in its current form, is that MothNet trains on and evaluates samples via the time-evolution of systems of coupled differential equations. 
This is time-consuming ($\sim$4 seconds per sample on a laptop), and would increase for more complex data sets with high-dimensional feature spaces, since these require larger networks with more neurons per layer and thus more equations to evolve.
In addition, the time-evolution system does not conveniently mesh with other ML platforms such as Tensorflow.
Thus, a future project is to develop different methods of running MothNet-like architectures that bypass the computations of time evolution and mesh with other platforms, yet functionally preserve a Hebbian update mechanism.
  

\subsubsection*{Acknowledgements}
Our thanks to Blake Richards, who articulated these hypotheses and suggested these experiments.\newline
CBD's work was partially supported by the Swartz Foundation.\newline 
JNK acknowledges support from the Air Force Office of Scientific Research (FA9550-19-1-0011).

 \bibliographystyle{icml2019.bst}
\bibliography{mothBibliography_june2019}

\end{document}